\documentclass[twocolumn,showpacs,fleqn,nobibnotes]{revtex4}

\usepackage{amsmath}
\usepackage{graphicx}
\usepackage{float}
\usepackage{subfigure}
\newcommand{\be}{\begin{equation}}
\newcommand{\ee}{\end{equation}}
\newcommand{\bea}{\begin{eqnarray}}
\newcommand{\ena}{\end{eqnarray}}

\begin{document}

\title{A study on the low mass Drell-Yan production at the CERN LHC within the dipole formalism}
\author{M.B. Gay Ducati, M.T. Griep and M.V.T. Machado}

\affiliation{High Energy Physics Phenomenology Group, GFPAE  IF-UFRGS \\
Caixa Postal 15051, CEP 91501-970, Porto Alegre, RS, Brazil}

\begin{abstract}
We present results for the low mass Drell-Yan production in proton-proton collisions at the LHC in the color dipole formalism. The DY differential cross sections  at $\sqrt{s}=7$ TeV as a function of dilepton rapidity, transverse momentum and invariant mass are discussed. We have imposed kinematical cuts  related to the low mass DY production investigated by ATLAS and LHCb collaborations. 

\end{abstract}

\maketitle
\section{Introduction}
\label{intro}

 The cross sections for producing lepton pairs by Drell-Yan (DY) process have been proven to still fulfill the factorization property and are finite to first  orders in perturbation theory at sufficiently large transverse momenta, $p_T$. On the other hand, there is an extensive program of research treating the low $p_T$ region as the conventional factorization approaches give divergent results at $p_T\rightarrow 0$ (see Ref. \cite{Kang} and references therein). In particular, in Ref. \cite{Klasen1} it was shown that differential cross section in the region $p_T\geq M_{\ell\ell}/2$ is driven by subprocesses initiated by incident gluons and therefore massive lepton-pair differential cross sections are useful sources of constraints on the gluon density. That study using next-to-leading order QCD and $p_T$-resummation was generalized in Ref. \cite{Klasen2} to polarized scattering and in Ref. \cite{Klasen3} to electroweak gauge boson production. The study of DY cross section with dileptons carrying large values of transverse momentum has long history and it is related to deep-inelastic-lepton scattering (DIS), prompt photon production and jet production as an important probe of
short-distance hadron dynamics. Besides helping to constraint the parton distribution functions (PDFs) in the nucleons, the great appeal of the production of dileptons in DY process in nuclear targets is that they are colorless probes of the dynamics of quarks and gluons \cite{Ayala}. Namely, they escape unscathed through
the colored medium of the high-energy collision. Thus, the dileptons can be a powerful probe of the initial state of matter created in heavy ion collisions,
since they interact with the medium only electromagnetically and
therefore provide a baseline for the interpretation of jet-quenching
models. Along these lines, it has been shown \cite{betemps,jamal_tam} that those electromagnetic probes are crucial to determine the dominant physics in the forward region at RHIC and at the LHC. 

 It has been shown \cite{Kope_gamma} that both direct (prompt) photon production and Drell-Yan dilepton pair production processes can be described within the same color dipole approach without any free parameters. Such a formalism, developed in \cite{bb} for the case of the total and diffractive cross sections, can be also applied to
radiation \cite{hir,brodsky}. In the rest frame of a target, the DY process looks like a bremsstrahlung \cite{Kopeliovich:2000fb}  of a massive photon from an incoming quark. The photons can be emitted before or after a quark to be scattered on a proton. Although in the process of electromagnetic
bremsstrahlung by a quark no real quark dipole participates, the cross section
can be expressed through the more elementary cross section
$\sigma_{dip}$ of interaction of a $Q\bar{Q}$ dipole \cite{Kopeliovich:2000fb} .  The relation between this formalism and the usual collinear pQCD factorization has been studied in details in Ref. \cite{Raufeisen:2002zp}.  The dipole formalism offers an easy way to calculate the transverse momentum distribution in DY process even in the low-$p_T$ region. The corresponding phenomenology investigating the role played by high energy approaches in the DY $p_T$ distribution has been investigated long before in Ref. \cite{Betemps:2003je}. Results from order $\alpha_s$ in the parton model cannot be directly compared to such approach since it is not an expansion in any parameter. All contribution from higher orders graphs enhanced by a factor $\ln (1/x_2)$ and even non-perturbative corrections are included. For instance, in  Ref. \cite{Golec} a twist expansion in powers of $\left(\frac{Q_{\mathrm{sat}}}{M_{\ell\ell}}\right)^2$ ($Q_{\mathrm{sat}}$ is the saturation scale and $M_{\ell\ell}$ is the invariant dilepton mass) was developed. It was shown that the leading twist is a good approximation to the full result for masses $M_{\ell\ell}\geq 6$ GeV. Recently, the diffractive DY cross section has been investigated in detail within the dipole framework \cite{Roman}. The corresponding factorization breaking effects in diffractive DY lead to very distinct properties of the observables compared to QCD factorisation-based calculations.

Our goal is to investigate in detail the low mass DY cross section at the LHC energies using color dipole approach, discussing several phenomenological aspects.  We focus mainly on forward rapidities and at the energy available at the LHC. The paper is organized as follows. In next section we summarize the main formula for DY differential cross section within the dipole framework, which is suitable for forward rapidities and allows to incorporate parton saturation effects. In last section we show our numerical results and predictions and summarize our main conclusions.

\section{Theoretical framework}

This section resumes the theoretical treatment for Drell-Yan production in $pp$ collisions in high energies considered in our study. We will use the dipole approach, which  is well suited for high-energy processes, i.e. small parton momentum fraction in the target $x_{2}\propto M_{\ell\ell}/\sqrt{s}$, and its range of validity is expected to be near $x_{2}<0.01$. The low mass DY production surely probes the small-$x$ physics, specially the forward rapidity case. One advantage in such an approach is to describe simultaneously the direct photon and dilepton production in the same theoretical framework and it provides finite cross sections in the limit $p_T\rightarrow 0$.  The transverse momentum $p_{T}$ distribution of virtual photon
bremsstrahlung in quark-nucleon interactions, integrated over the
final quark transverse momentum, was derived in \cite{Kopeliovich:2000fb} in terms
of the dipole formalism,
 \begin{eqnarray}
\frac{d^3 \sigma^{qN}(q\to q\gamma^*)}{d(ln \alpha)\,d^{2}\vec{p}_{T}} & = & \frac{1}{(2\pi)^{2}}
\sum_{in,f}\sum_{L,T}
\int d^{2}\vec{r}_{1}d^{2}\vec{r}_{2}e^{i \vec{p}_{T}.(\vec{r}_{1}-\vec{r}_{2})} \nonumber \\
&\times & \phi^{\star T,L}_{\gamma q}(\alpha, \vec{r}_{1})
\phi^{T,L}_{\gamma q}(\alpha, \vec{r}_{2}) \nonumber \\
&\times & \left[ \frac{1}{2}\left(
\sigma_{dip}(x_2,\alpha r_{1})+\sigma_{dip}(x_2,\alpha r_{2})\right) \right. \nonumber \\
&- & \left. \frac{1}{2}\sigma_{dip}(x_2,\alpha(\vec{r}_{1}-\vec{r}_{2})) \right], 
\label{di}
\end{eqnarray}
where $\vec{r}_{1}$ and $\vec{r}_{2}$ are the quark-photon transverse
 separations in the two radiation amplitudes contributing to the cross
 section, $\sigma_{dip}$. The parameter $\alpha$ is the
 relative fraction of the quark momentum carried by the photon, and is
 the same in both amplitudes, since the interaction does not change
 the sharing of longitudinal momentum.  In the equation above, $T$
 stands for transverse and $L$ for longitudinal photons. The energy
 dependence of the dipole cross section, which comes through the variable
 $x_2=2 (p_1\cdot q)/s=(M_T/\sqrt{s})\,e^{-y}$ (with $M_T=\sqrt{M^2+p_T^2}$), where $p_1$ is the projectile four-momentum and $ q$ is
the four-momentum of the dilepton, is generated by additional
radiation of gluons which can be resummed in the leading $\ln(1/x)$
approximation.

In Eq.~(\ref{di}) the light-cone wavefunction of the projectile
quark $\gamma q$ fluctuation has been decomposed into  transverse
$\phi^{T}_{\gamma q}(\alpha, \vec{r})$ and longitudinal
$\phi^{L}_{\gamma q}(\alpha, \vec{r})$ components,  and an average
over the initial quark polarization and sum over all final
polarization states of quark and photon is performed. The expressions for $T$ and $L$ wavefunction components are well known at the lowest order \cite{Kopeliovich:2000fb,Betemps:2003je}. The hadron cross section can be obtained from the elementary
partonic cross section Eq.~(\ref{di}) summing up the
contributions from quarks and antiquarks weighted with the
corresponding parton distribution functions (PDFs),
\begin{eqnarray}
\frac{d^4 \sigma (pp\to \ell^+\ell^- X)}{dydM^2d^{2}\vec{p}_{T}}& = & K_{\mathrm{eff}}\frac{\alpha_{em}}{3\pi M^2}\int_{x_{1}}^{1}\frac{d\alpha}{\alpha} F_{2}^{p}\left(\frac{x_{1}}{\alpha},\mu^2\right)\nonumber \\
&\times & \frac{d \sigma^{qN}(q\to q\gamma^*)}{d(ln \alpha)\,d^{2}\vec{p}_{T}},\
\label{con}
\end{eqnarray}
where the PDFs of the projectile have entered in a combination which can be
written in terms of proton structure function $F_{2}^{p}$, with $x_1=(M_T/\sqrt{s})\,e^{y}$. For the hard scale $\mu$ entering into the proton
structure function in Eq.~({\ref{con}}), we take $\mu^{2}=\beta[(1-x_1)M^2+p^{2}_{T}]$ and the energy scale of the dipole cross section in Eq.~(\ref{di}) is given by the variable $x_{2}$. The dependence of the cross section on the choice for the hard scale can be obtained by varying the $\beta$ value (the default value here is $\beta=1$). The quantity $K_{\mathrm{eff}}$ takes into account the effective higher-order DY contributions. For simplicity we take the expression \cite{Barger}:
\begin{eqnarray}
K_{\mathrm{eff}} (\mu^2) = 1+ \frac{\alpha_s(\mu^2)}{2\pi}\left(1+ \frac{4}{3}\pi^2  \right),
\end{eqnarray}
where the running coupling constant is computed  at the scale $\mu^2=M^2$.

An important piece in the color dipole calculations is the dipole cross section. It is theoretically unknown, although several parameterizations have been proposed. Here, we consider some analytical parameterizations which rely on the geometric scaling property. In this case, they are a function of a scaling variable $rQ_{sat}(x)$ where $Q_{sat}$ is the so called saturation scale. It defines the transverse momentum scale where parton recombination physics is relevant and in general is modeled as $Q_{sat}\propto x^{-\lambda/2}$. A common feature on these models is that for decreasing $x$, the
dipole cross section saturates for smaller dipole sizes. In addition, at
small $r$, as perturbative QCD implies $\sigma\sim r^{2}$, they  vanish, i.e. the color transparency phenomenon occurs. In a general form, they can be parameterized as \cite{gbw},
\begin{equation}
\sigma_{dip}(x,\vec{r};\gamma)=\sigma_{0}\left[ 1-\exp\left(-\frac{r^{2}Q_{sat}^{2}}{4}\right)^{\gamma_{\mathrm{eff}}}\,\right],\label{gbw}
\end{equation}
where the quantity $\gamma_{\mathrm{eff}}$ is the effective anomalous dimension. The GBW parameterization \cite{gbw} uses  $\gamma_{\mathrm{eff}} = 1$ and the remaining parameters are fitted to DIS HERA data at small $x$. The saturation scale is defined as $Q_{sat}^2(x)=\left(\frac{x_0}{x}\right)^{\lambda}$. This
parameterization gives a quite good description of DIS data at $x<10^{-2}$.

The main difference among the distinct phenomenological models using parameterizations as Eq. (\ref{gbw}) comes from the  predicted behavior for the anomalous dimension, which determines  the  transition from the non-linear to the extended geometric scaling regime, as well as from the extended geometric scaling to the DGLAP regime. It is  the behavior of $\gamma$ that determines the diminishing of the hadronic cross section as $p_T$ increases. Several models in the literature consider the general form $\gamma_{\mathrm{eff}} = \gamma_{sat} + \Delta (x,r;p_T)$, where $\gamma_{sat}$ is the anomalous dimension at the saturation scale and $\Delta$ mimics the onset of the geometric scaling region and DGLAP regime. In order to take into account this possibility, here we also will  consider the  phenomenological saturation model proposed in Ref. \cite{iim} which encodes the
main properties of the saturation approaches, with the dipole cross section  parameterized as follows
\begin{eqnarray}
\sigma_{dip}\,(x,r) =\sigma_0\,\left\{ \begin{array}{ll} 
{\mathcal N}_0 \left(\frac{\bar{\tau}^2}{4}\right)^{\gamma_{\mathrm{eff}}\,(x,\,r)}\,, & \mbox{for $\bar{\tau} \le 2$}\,,  \nonumber \\
 1 - \exp \left[ -a\,\ln^2\,(b\,\bar{\tau}) \right]\,,  & \mbox{for $\bar{\tau}  > 2$}\,, 
\end{array} \right.
\label{CGCfit}
\end{eqnarray}
where $\bar{\tau}=r Q_{\mathrm{sat}}(x)$ and the expression for $\bar{\tau} > 2$  (saturation region)   has the correct functional
form, as obtained  from the theory of the Color Glass Condensate (CGC) \cite{iim}. For the color transparency region near saturation border ($\bar{\tau} \le 2$), the behavior is driven by the effective anomalous dimension $\gamma_{\mathrm{eff}}\, (x,\,r)= \gamma_{\mathrm{sat}} + \frac{\ln (2/\tilde{\tau})}{\kappa \,\lambda \,y}$, where $\gamma_{\mathrm{sat}}=0.63$ is the LO BFKL anomalous dimension at saturation limit.

Before computing numerically the cross section given by Eq. (\ref{con}), we discuss about the semi-analytical calculation which is allowed in color dipole picture in the color transparency region. In that case, an expression for $p_T$ distribution can be written using Eq. (\ref{con}) and the expressions for the transverse momentum $p_{T}$ distribution of photon
bremsstrahlung in quark-nucleon interactions, Eq. (\ref{di}). The explicit equation for the DY differential cross section, Eq. (\ref{con}), reads as:
\begin{eqnarray}
& & \frac{d^4\sigma\,(pp\to \ell^+\ell^-
X)}{dydM^2 d^{2}\vec{p}_{T}}  = 
\frac{\alpha_{em}^2}{6\pi^3M^2}\int_{x_{1}}^{1}\frac{d\alpha}{\alpha}
 F_{2}^{p}\left(\frac{x_{1}}{\alpha},Q^2=\mu^2\right) \nonumber\\
&\times & \left \{ \left[ m_q^2\alpha^4+2M^2(1-\alpha)^2\right]\left[\frac{{\cal I}_1}{(p_T^2+\varepsilon^2)}-\frac{ {\cal I}_2}{4\varepsilon} \right] \right. \nonumber \\
& + & \left.  [1+(1-\alpha)^2]\left[ \frac{\varepsilon p_T \, {\cal I}_3}{(p_T^2+\varepsilon^2)} -\frac{{\cal I}_1}{2}+\frac{\varepsilon \,{\cal I}_2}{4}\right]
\right \},
\end{eqnarray}
 where $\varepsilon^2 = (1-\alpha)M^2+\alpha^2 \,m_q^2$. The quantities ${\cal I}_{1,2,3}$ are Hankel's integral transforms of order $0$ (${\cal I}_{1,2}$) and order $1$ (${\cal I}_{3}$) given by:
\begin{eqnarray}
{\cal I}_1 & = & \int_0^{\infty}dr\,rJ_0(p_T\,r)K_0(\varepsilon\,r)\, \sigma_{dip}(x_2,\alpha r), \nonumber \\
{\cal I}_2  & = &  \int_0^{\infty}dr\,r^2J_0(p_T\,r)K_1(\varepsilon\,r)\, \sigma_{dip}(x_2,\alpha r),\nonumber \\
{\cal I}_3 & = & \int_0^{\infty}dr\,rJ_1(p_T\,r)K_1(\varepsilon\,r)\, \sigma_{dip}(x_2,\alpha r).
\label{ints}
\end{eqnarray}
Here,  $K_{0,1}(x)$ denotes the
modified Bessel function of the second kind and  $m_{q}$ is an effective quark mass. 

Considering a kinematic interval where the dipole cross section is dominated by the color transparency region, one can use GBW parameterization and take its small-$r$ limit to compute analytically the integrals in Eq. (\ref{ints}).  In this case, we can take the approximation $\sigma_{dip}\approx \sigma_0(r^2\,Q_{sat}^2)$  in the region where $p_T\gg Q_{sat}$. At the LHC energies, the typical saturation scale is of units of GeV. For instance, at mid-rapidity $x_2\simeq M/\sqrt{s}$, which gives for $\langle M  \rangle = 10 $ GeV a value $x_2\approx 10^{-3}$ at 7 TeV. Then, the saturation scale is of order $Q_{\mathrm{sat}}=(x_0/x_2)^{\lambda/2}\,\mathrm{GeV}\approx 0.8$ GeV (with $x_0=3.04\times 10^{-4}$ and $\lambda = 0.288$).  The final results for the hadron cross section is then,
\begin{eqnarray}
& & \frac{d \sigma(pp\to \ell^+\ell^-X)}{dydM^2d^{2}\vec{p}_{T}}  \approx 
\frac{\alpha_{em}^2\sigma_0Q_{sat}^2}{6\pi^3M^2}\int_{x_{1}}^{1}\frac{d\alpha}{\alpha}
 F_{2}^{p}\left(\frac{x_{1}}{\alpha},Q^2\right)\nonumber \\
&\times &\left \{ \left[ m_q^2\alpha^4+2M^2(1-\alpha)^2\right]\left[ \frac{p_T^2}{(p_T^2+\varepsilon^2)^4}\right]  \right.\nonumber \\
& + &  \left. [1+(1-\alpha)^2]\left[ \frac{p_T^4+\varepsilon^4}{2(p_T^2+\varepsilon^2)^4}\right] \right \} 
\label{analitic}
\end{eqnarray}

In what follows we compute the differential cross sections of low mass DY production in hadron-hadron collisions focusing in the recent measurements done by ATLAS  \cite{ATLAS} and LHCb \cite{LHCb} collaborations at the LHC.

\section{Numerical results}

\begin{figure}[t]
\includegraphics[scale=0.4]{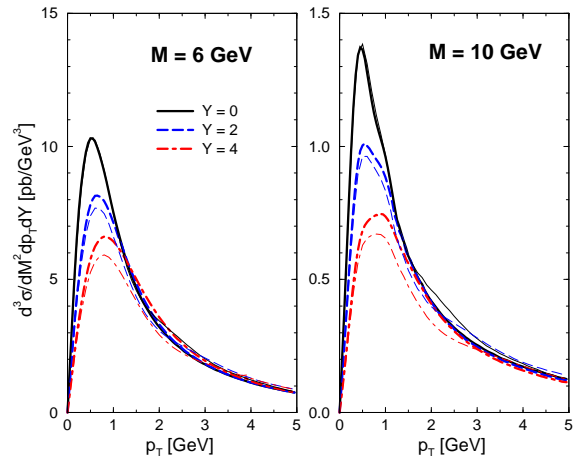}
\caption{Low mass DY differential cross sections, $d^3\sigma /dM^2dYdp_T$, as a function of dilepton transverse momentum, $p_T$, at energy of $\sqrt{s}=7$ TeV. The plots are shown for fixed dilepton mass ($M=6$ and $10$ GeV) and distinct lepton pair rapidities ($Y=0,2,4$). The results are presented using the GBW dipole cross section (bold curves) and the CGC dipole cross section (thin curves).}
\label{fig:1}
\end{figure}

Let us now present some numerical calculations concerning the low mass dilepton production in the LHC energy regime of $\sqrt{s}=7$ TeV. For the proton structure function in Eqs.~(\ref{con}) we have taken the ALLM parameterization \cite{ALLMnew}, which is valid in the kinematic range we are interested in. The sensitivity to a different choice for $F_2$ is very small. Moreover, in order to account for the threshold region $x_2\rightarrow 1$, we have corrected the dipole cross section by multiplying it by a threshold factor $(1-x_2)^7$. We consider $m_q=0.2$ GeV for the effective quark mass. In Fig.~\ref{fig:1}, we show the results for the differential cross section,  $d^3\sigma /dM^2dYdp_T$ (in units of pb), as a function of the dilepton transverse momentum $p_T$. Here, the predictions are obtained using the GBW dipole cross section (bold curves) and the CGC dipole cross section (thin curves) and using the hard scale $\mu^2 = (1-x_1)M^2+p_T^2$. Notice that the $p_T$-spectrum is quite sensitive to the particular model of dipole cross section (specially at large transverse momentum) as it depends on the behavior of effective anomalous dimension as discussed in the previous section. In the left panel is shown the case for fixed invariant mass $M=6$ GeV and for sample values of dilepton rapidity including central and forward rapidities, i.e. $Y=0,2$ and 4, respectively. The same notation holds for the right panel, where now the invariant mass is $M=10$ GeV. As expected, the large rapidity cases give smaller cross sections and the peak on the distributions is shifted to larger values of transverse momentum. In the kinematical situation investigated here the peak is located at momentum around $p_T\approx 1$ GeV. The shift and location of the peak can be understood looking at the semi-analytical expression, Eq. (\ref{analitic}). A clear problem in the numerical calculations using the full $p_T$ spectrum is due to the strongly oscillating integrand appearing in Eqs. (\ref{ints}), which turn out the $p_T$ integrations a delicate task mainly for high mass values.

\begin{figure}[t]
\includegraphics[scale=0.4]{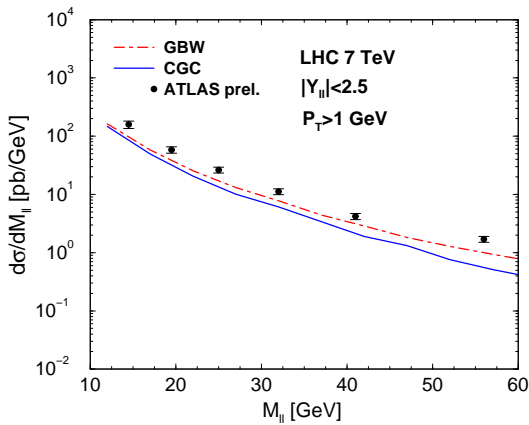}
\caption{Invariant mass distribution in the range $12< M_{\ell\ell}< 60$ GeV. The imposed cuts at energy of $\sqrt{s}=7$ TeV are lepton pair rapidities $|Y_{\ell\ell}|<2.5$ and dilepton transverse momentum $p_T\geq 1$ GeV. Preliminary ATLAS data  \cite{ATLAS} are shown for sake of comparison.}
\label{fig:2}
\end{figure}
 
In Fig.~\ref{fig:2} we show the invariant mass distribution at midrapidities obtained from two different implementations of the dipole cross section taken from recent phenomenological works. We considered the GBW model (dot-dashed line) and  the  phenomenological saturation model, labeled here CGC (solid line), which involves a running anomalous dimension.  The main deviation between these two models occurs at large $p_T$, which gives distinct overall normalizations for the dilepton invariant mass distribution. The considered cuts are along the lines presented by the ATLAS analysis \cite{ATLAS} for low mass Drell-Yan di-muon process. The selection cuts on that analysis at energy of $\sqrt{s}=7$ TeV and integrated luminosity of 36 pb$^{-1}$ were low muon transverse momentum, $p^{\mu}_T>6$ GeV and low di-muon mass region $12< M_{\ell\ell}< 66$ GeV. Here, we consider the integration over the boson rapidity in the range $|Y_{\ell\ell}|<2.5$ and dilepton transverse momentum $p_T\geq 1$ GeV. Distinct $p_T$ cuts will lead to a different overall normalization for the invariant mass distribution. At this stage we did not impose the selected cuts on individual muons as done by ATLAS analysis. Two different studies were done by ATLAS \cite{ATLAS}: selection of muons of different minimum $p^{\mu}_T$ and no requirement on boson selection except for the invariant mass constraint (asymmetric analysis) and selection of minimum $p^{\mu}_T$ for muons and a requeriment on the rapidity of boson is applied (symmetric analysis). The results presented here are somewhat consistent with the extrapolated Born level differential cross section using the symmetric analysis. For sake of comparison, we include the preliminary data \cite{ATLAS} in Fig. \ref{fig:2} (filled circles). 

The main $M^2$ dependence of DY cross section can be quantitatively understood in the color dipole framework. In the case of integrated cross section on $p_T$, it was shown in Ref. \cite{Golec} that a twist expansion in positive powers of the ratio $\left(\frac{Q_{\mathrm{sat}}}{M_{\ell\ell}}\right)^2$  can be performed. The leading twist contribution reads as \cite{Golec}:
\begin{eqnarray}
\frac{d^2\sigma}{dydM^2} & \approx & \frac{\alpha_{em}^2\sigma_0}{12\pi^2M^2} \frac{Q_{\mathrm{sat}}^2(x_2)}{M^2}
 F_{2}^{p}\left(x_{1},M^2\right)\nonumber \\
&\times & \left\{\frac{4}{3} \gamma_E-1 +\frac{2}{3}\left[ \psi\left(\frac{5}{2}\right) - \ln \frac{Q_{\mathrm{sat}}^2(x_2)}{4M^2(1-x_1)}\right]\right\}, \nonumber  
\label{Golec}
\end{eqnarray}
modulo contributions of ${\cal O}(1-x_1)$. The exact leading twist formula is quite close to the all-twist result at the LHC energies in the region of $M>6$ GeV \cite{Golec}. In central rapidities considered above, $\langle Y  \rangle =0$, the saturation scale changes from $\langle Q_{\mathrm{sat}}^2 \rangle \simeq 0.4$ GeV$^2$  down to $\langle Q_{\mathrm{sat}}^2 \rangle \simeq 0.2$ GeV$^2$   in the invariant mass range $12< M_{\ell\ell}< 66$ GeV, respectively. In this situation, the analytical result given in  Eq. (\ref{analitic}) should be quite reliable. It would be timely in the future to investigate the effects of QCD DGLAP evolution in the dipole cross section and also the impact parameter dependence in the CGC model as done recently in Refs. \cite{Reza1} and  \cite{Reza2}.

\begin{figure}[t]
\includegraphics[scale=0.4]{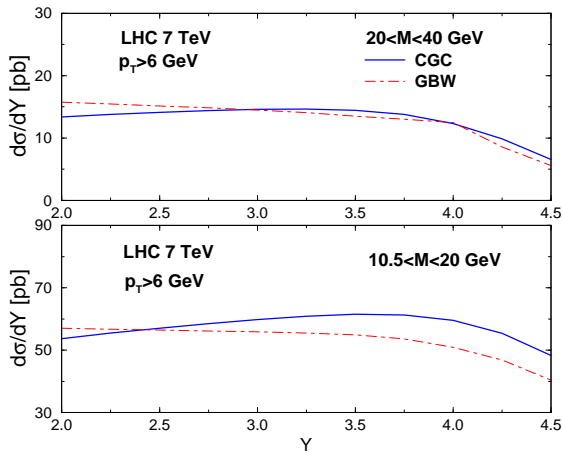}
\caption{The dilepton rapidity distribution at $\sqrt{s}=7$ TeV imposing the cut on dimuon transverse momentum $p_T>6$ GeV and two invariant mass regions: (upper plot)  $20\leq M_{\ell\ell}\leq 40$ GeV and (lower plot) $10.5\leq M_{\ell\ell}\leq 20$ GeV.}
\label{fig:3}
\end{figure}

Now, we focus on the forward rapidity region. In Fig.~\ref{fig:3}, the rapidity distribution, $d\sigma/dY$, is computed for the interval $2<Y<4.5$ considering the two phenomenological models (same notation as the previous plot) and using the hard scale $\mu^2 = \frac{1}{2}[(1-x_1)M^2+p_T^2]$. It was found that the rapidity distribution is sensitive to the chosen hard scale as also occurs for the LO pQCD approach. Here, the deviations comparing the models are stronger than for the mass distribution. The main point is that the rapidity distribution is driven by the effective anomalous dimension, which is distinct in the models. We have imposed a cut for the dilepton transverse momentum of $p_T>6$ GeV and two distinct intervals of invariant mass.  Namely, in the upper plot one has  $20\leq M_{\ell\ell}\leq 40$ GeV whereas in the lower plot one has $10.5\leq M_{\ell\ell}\leq 20$ GeV. The motivation for such a cut is due to the recent LHCb collaboration \cite{LHCb} measurement of low mass DY cross section. The measurement collected with an integrated luminosity of 37 pb$^{-1}$ are for the di-muon final state having muons within pseudorapidities of 2 to 4.5, muon transverse momentum $p^{\mu}_T>3$ GeV  ($p^{\mu}_T>15$ GeV for higher masses) in two distinct mass regions. In the forward rapidities considered here, the saturation scale is in the interval $0.6 \leq \langle Q_{\mathrm{sat}}^2 \rangle \leq 1.2$ GeV$^2$  for  $\langle M_{\ell\ell} \rangle \simeq 15.25$ GeV. Slightly lower  values are found also for higher mass $\langle M_{\ell\ell} \rangle \simeq 30$ GeV.

\begin{figure}[t]
\includegraphics[scale=0.35]{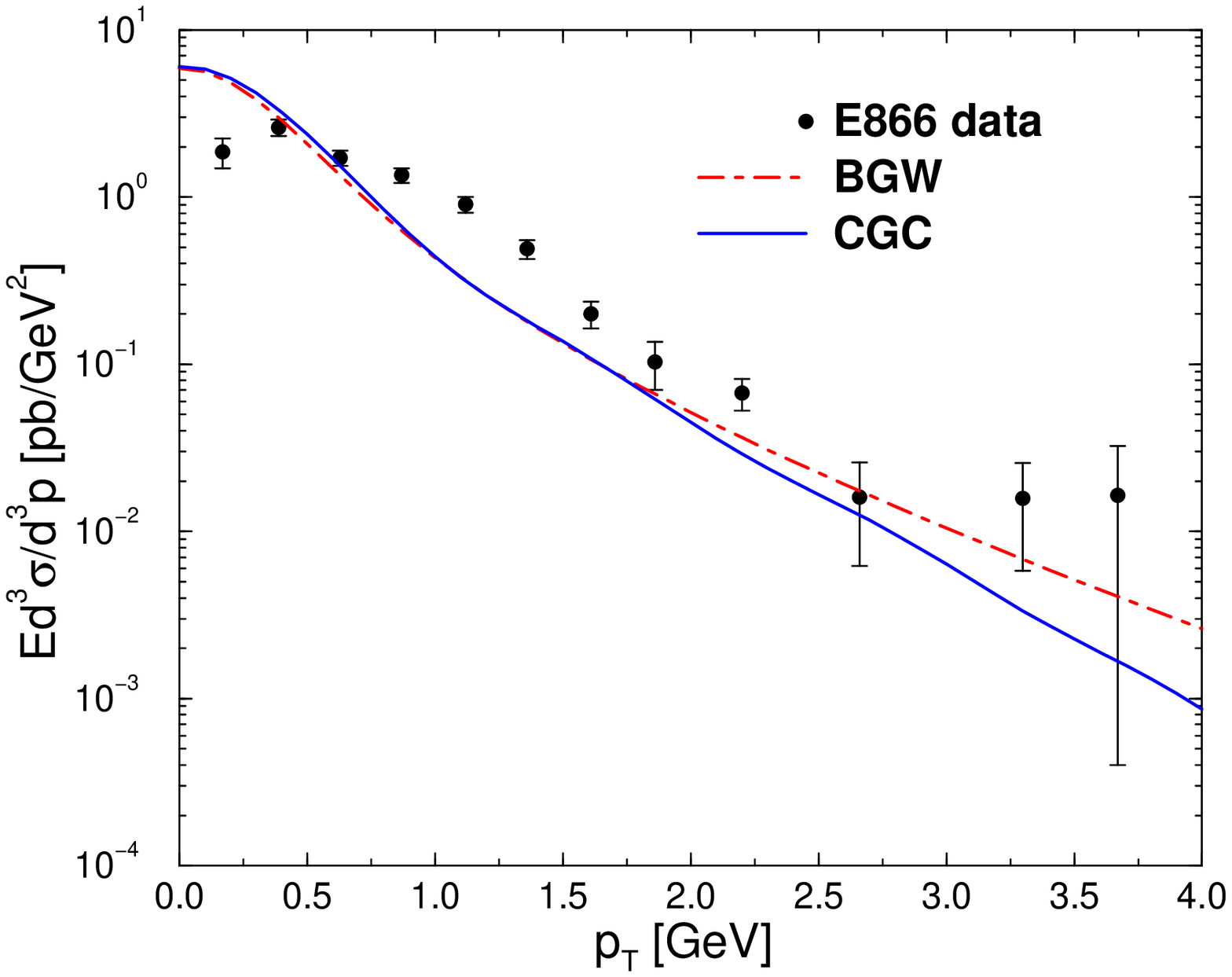}  \includegraphics[scale=0.35]{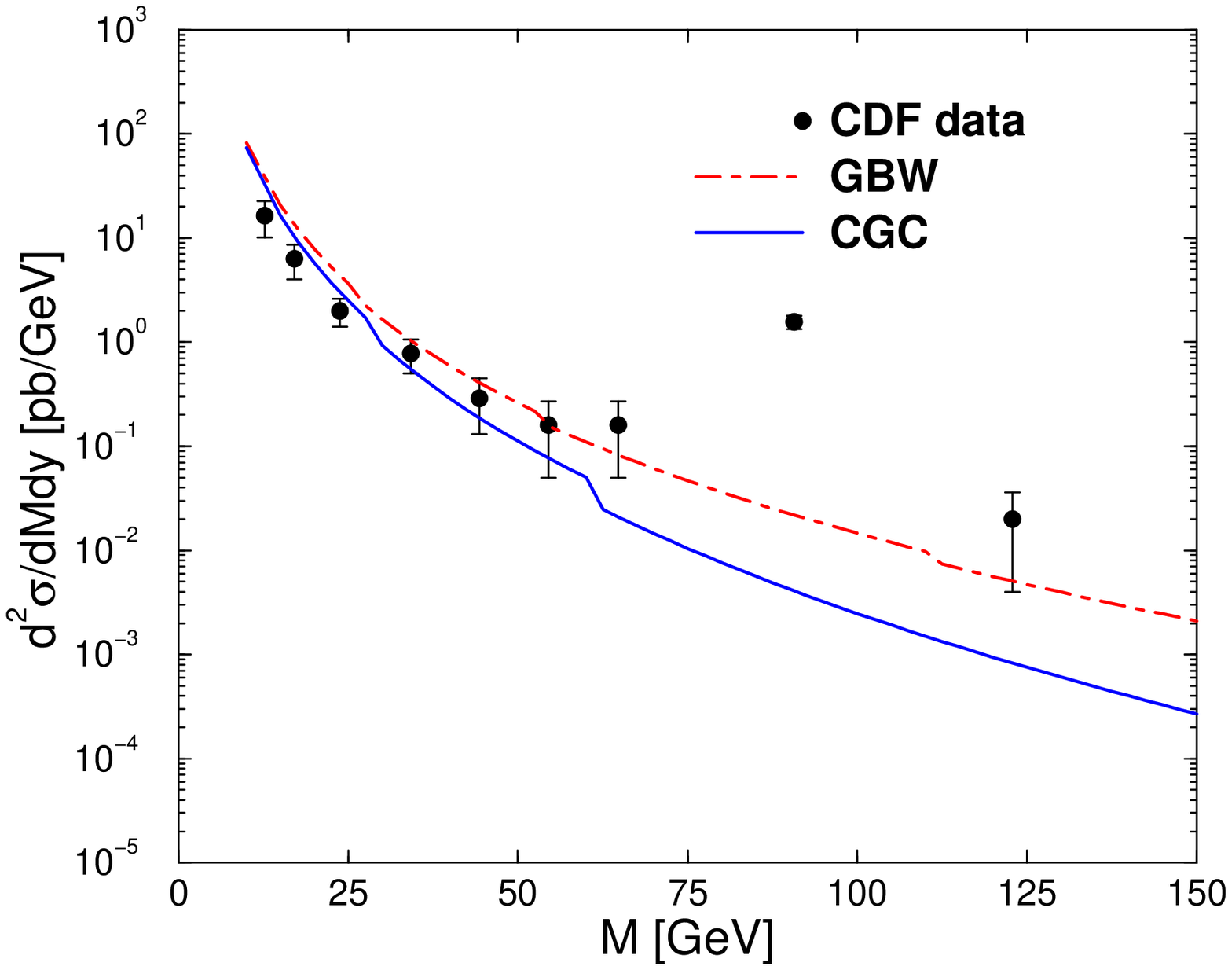}
\caption{The DY invariant cross section (upper panel) at $\sqrt{s}= 39$ GeV as a function of $p_T$ compared to the E866 Collaboration data \cite{E866}. In bottom panel, the differential cross section $d^2\sigma /dMdy$ at $\sqrt{s}=1800$ GeV as a function of invariant mass is presented and compared to the CDF Collaboration data \cite{CDFDY}.}
\label{fig:4}
\end{figure}

Finally, we check the energy dependence of the DY differential cross section within the color dipole picture. In Fig. \ref{fig:4} (upper panel) we show the invariant cross section as a function of $p_T$ at energy $\sqrt{s}= 39$ GeV. The experimental results from the E866 Collaboration  \cite{E866} are also presented ($\langle x_F \rangle \simeq 0.63 $ and $4.2 \leq M_{\mu^+\mu^-}\leq 5.2$ GeV). In the bottom panel, the differential cross section $d^2\sigma / dMdy$ (for $|y|<1$) is shown for the energy $\sqrt{s}=1800$ GeV as a function of dilepton invariant mass. The data from CDF Collaboration \cite{CDFDY} are included in the plot, considering also the large invariant mass data points. The solid curves refer to CGC and dot-dashed curves to GBW dipole cross section, respectively. The color dipole picture reasonably describes the cross section from low to high energies in the kinematical regions where it is expected to be valid. The approach is also somewhat consistent with calculations carried out in next-to-leading order QCD at both fixed target and collider energies \cite{Klasen1}. 

As a summary, we showed that low mass DY  production can be addressed in the color dipole picture without any free parameters by using dipole cross section determined from current  phenomenology in DIS. It has been shown before \cite{Mariotto} that in central rapidities at RHIC and Tevatron, saturation effects do not play a significant role for the measured range of $p_{T}$. This situation can be changed at the LHC even at midrapidities as the saturation scale is enhanced by a sizable factor.  We write down also analytical results relevant for the calculation of $p_T$ spectrum of DY dileptons. 

\section*{Acknowledgments}
This work was  partially financed by the Brazilian funding
agencies CNPq and FAPERGS and by the French-Brazilian scientific cooperation project CAPES-COFECUB 744/12.  MVTM thanks Murilo Rangel, Ronan McNulty and Michael Klasen for helpful comments and MBGD thanks M.A. Betemps for useful discussions.

\end{document}